\documentclass[astronomy,article,accept,pdftex,moreauthors]{Definitions/mdpi} 
\firstpage{1} 
\pubvolume{1}
\issuenum{1}
\articlenumber{0}
\pubyear{2026}
\copyrightyear{2026}
\externaleditor{Firstname Lastname} 
\datereceived{4 June 2026} 
\daterevised{5 July 2026} 
\dateaccepted{24 July 2026} 
\datepublished{ } 
\usepackage{comment}
\usepackage{rotating}
\usepackage{float}

\Title{Radio Jet Properties of the Flaring Gamma-Ray Blazar PKS~0346$-$27}

\Author{M\'at\'e Krezinger $^{1,2}$\orcidA{}, S\'andor Frey $^{1,2,3,}$*\orcidB{} and Krisztina \'E. Gab\'anyi $^{1,2,3,4}$\orcidC{}}

\AuthorNames{M\'at\'e Krezinger, S\'andor Frey, and Krisztina \'E. Gab\'anyi}

\address{%
$^{1}$ \quad Konkoly Observatory, HUN-REN Research Centre for Astronomy and Earth Sciences, Konkoly Thege Mikl\'{o}s \'{u}t 15-17, 1121 Budapest, Hungary; krezinger.mate@csfk.org (M.K.); k.gabanyi@astro.elte.hu (K.\'E.G.) \\
$^{2}$ \quad Research Centre for Astronomy and Earth Sciences (CSFK), MTA Centre of Excellence, Konkoly Thege Mikl\'{o}s \'{u}t 15-17, 1121 Budapest, Hungary\\
$^{3}$ \quad Department of Astronomy, Institute of Physics and Astronomy, ELTE E\"otv\"os Lor\'and University, \mbox{P\'azm\'any P\'eter s\'et\'any 1/A,} 1117 Budapest, Hungary\\

$^{4}$ \quad HUN-REN--ELTE Extragalactic Astrophysics Research Group, ELTE E\"otv\"os Lor\'and University, \mbox{P\'azm\'any P\'eter s\'et\'any 1/A,} 1117 Budapest, Hungary\\
}

\corres{Correspondence: frey.sandor@csfk.org}

\abstract{
The blazar PKS~0346$-$27 at redshift $z=0.991$ is among the most distant extragalactic sources detected in very high energy $\gamma$-rays. It underwent a major flare reaching TeV energies in November 2021. Recent modelling of its spectral energy distribution favoured a single-zone proton--synchrotron-dominated hadronic model with bulk Lorentz factor $\Gamma=10$ over a single-zone leptonic jet model with much higher $\Gamma$. Here we analyse archival radio data to give independent estimates of the jet bulk Lorentz factor. We present high-resolution radio imaging and brightness distribution modelling, as well as flare detection in long-term radio flux density monitoring data, and obtain $\Gamma \lesssim 10$ values. This lends support to the hadronic model proposed for the PKS~0346$-$27 jet. Our results highlight the importance of long-term, very long baseline interferometry and total flux density monitoring in studying blazar jets.
}


\begin{document}

\section{Introduction}

Blazars are a class of active galactic nuclei (AGN) that harbour powerful relativistic jets aligned close to the line of sight, i.e., within $1/\Gamma$, where $\Gamma$ is the bulk Lorentz factor of the jet \cite{1995PASP..107..803U}. For $\Gamma \gtrsim 5$, the viewing angles of blazars are $<$$10^{\circ}$. Relativistic effects cause the blazar jet emission Doppler-boosted, making the enhanced emission detectable even from large cosmological distances. Blazars are characterized by large and rapid brightness variations, from the radio to the $\gamma$-ray regime (e.g., \cite{2025A&ARv..33....8R}).

The broadband spectral energy distributions (SEDs) of blazars show a double-peaked profile, peaking in the infrared (IR) to X-ray for the low-energy component, and in the MeV to TeV bands for the high-energy component. The low-energy component originates from the synchrotron radiation of the relativistic electrons in the jet, whereas the high-energy component is often explained by leptonic models, like inverse-Compton (IC) process by electrons directly accelerated within the jet (e.g.,~\cite{2006IJMPA..21.6015L,2007Ap&SS.309...95B}). In hadronic models, the high-energy emission can originate from both electron and proton--synchrotron radiation, or the decay products following photo-pion interactions of relativistic protons (e.g.,~\cite{2013ApJ...768...54B}). Steady-state leptonic models have been applied with great success in modelling the SEDs of \textit{Fermi} $\gamma$-ray blazars (e.g.,~\cite{2008MNRAS.385..283C,2010MNRAS.402..497G}), while time-dependent models that explore the variability features are also making progress (e.g., \cite{2011ApJ...727...21J}). There are still issues with the very fast (i.e., time scales of a few minutes) variability of some TeV blazars, since the single-zone models would require extremely high bulk Lorentz factors ($\Gamma \gtrsim 50$, \cite{2007ApJ...669..862A,2007ApJ...664L..71A,2008MNRAS.384L..19B}). To overcome the difficulties, multi-zone models have been proposed, including the spine--sheath \mbox{model \cite{2008MNRAS.385L..98T}}, a decelerating-jet model \cite{2003ApJ...594L..27G}, and internal-shock models (e.g.,~\cite{2008ApJ...689...68G,2011ApJ...727...21J}).

The bulk Lorentz factor is an intrinsic parameter of the jet, which describes the speed of its flow and can be calculated from the Doppler-boosting factor ($\delta$) and the apparent jet speed ($\beta_\mathrm{app}$) \cite{1995PASP..107..803U}. 
Here $\beta_\mathrm{app}$ can be measured with the high-resolution radio astronomical technique of very long baseline interferometry (VLBI), if long-term imaging data are available (e.g.,~\cite{2008A&A...484..119B,2012ApJ...758...84P,2017ApJ...846...98J,2019ApJ...874...43L,2021ApJ...923...30L}). The Doppler factor can be estimated in different ways \cite{1999ApJ...511..112L}. The simplest method is to directly measure the radio core brightness temperature with VLBI and compare it to the intrinsic brightness temperature. Variability Doppler factor can be estimated through total flux density variability data (e.g.,~\cite{1999ApJS..120...95V,2009A&A...494..527H,2010A&A...521A...6V}). Yet another way to estimate $\delta$ is to compare X-ray emission originating from inverse-Compton-scattered photons and radio emission produced by relativistic electrons in the jet \cite{1993ApJ...407...65G,1996ApJ...461..600G}. For this method, the knowledge of the radio flux density at the spectral turnover frequency is required. Estimating $\delta$ in different ways gives constraints to the value of the bulk Lorentz factor. This way, we are able to check the viability of the jet models mentioned above, and decide which one is more suitable to explain the SED of a given blazar.

PKS~0346$-$27 is a prominent blazar, more specifically a flat-spectrum radio quasar (FSRQ) already identified in the 1960s \cite{1964AuJPh..17..340B}. Its spectroscopic redshift is $z = 0.991$ \cite{1988ApJ...327..561W}. Its coordinates in the VLBI-defined International Celestial Reference Frame (ICRF3, \cite{2020A&A...644A.159C}) are right ascension $\mathrm{RA}=57.15893571^{\circ}$ and declination $\mathrm{Dec}=-27.82043494^{\circ}$. The source was detected in X-rays by the \textit{Roentgen Satellite} (\textit{ROSAT}) \cite{1999A&A...349..389V} and in $\gamma$-rays by the \textit{Fermi} Large Area Telescope (LAT) \cite{2010ApJS..188..405A}. It is associated with the $\gamma$-ray source 4FGL~J0348.5$-$2749 \cite{2020ApJS..247...33A}. PKS~0346$-$27 already showed multiple $\gamma$-ray flares, the first one detected on 2 February 2018 was reported by \cite{2018ATel11251....1A}. This flare was immediately followed up and confirmed in near-infrared \cite{2018ATel11269....1N}, then in ultraviolet and X-ray \cite{2018ATel11455....1N} wavebands. A long-term multi-wavelength variability study identified five more $\gamma$-ray flaring episodes from December 2019 to January 2021 \cite{2023MNRAS.520.2024K}. The minimum variability time scale is $\sim$$1$~day, suggesting the compactness of the emission region in the source. On 3 November 2021, Ref. \cite{2021ATel15020....1W} reported the first-time detection of very-high-energy (VHE, $E>100$~GeV) $\gamma$-ray photons from PKS~0346$-$27 by the \textit{Fermi}-LAT. It was followed by observations with the High Energy Stereoscopic System (H.E.S.S.) \cite{2024icrc.confE.924C,2026A&A...706A.246H}. H.E.S.S. detected the source $\sim$$2$ days after the peak of a prominent flare observed by \textit{Fermi}-LAT. Notably, PKS~0346$-$27 is the second most distant VHE $\gamma$-ray source currently known, closely following OP\,313 at $z = 0.997$ \cite{2023ATel16381....1C}.

In this paper, we present the radio analysis of PKS~0346$-$27 during the time of the reported TeV flaring event. By estimating the jet bulk Lorentz factor using two different methods based on radio light curve variability and high-resolution VLBI imaging of PKS~0346$-$27, we are able to compare the results with the values estimated recently from SED modelling \cite{2026A&A...706A.246H}. Throughout this paper, we assume a standard flat $\Lambda$ Cold Dark Matter cosmological model with $\Omega_{\mathrm m} = 0.3$, $\Omega_{\Lambda} = 0.7$, and $H_0 = 70$~km\,s$^{-1}$\,Mpc$^{-1}$. With these parameters, the luminosity distance of PKS~0346$-$27 at $z = 0.991$ is $6533.6$~Mpc \cite{2006PASP..118.1711W}. 

\section{Observations and Data Reduction}

\subsection{Archival Data} 

PKS~0346$-$27 is used as a calibrator radio source in the southern sky, and its flux density has been monitored regularly in five different radio bands, namely $L$ (1.1--3.1 GHz), $C$--$X$ (4--11 GHz), $U$ (16--26 GHz), and $Q$ (30--50 GHz), at the Australia Telescope Compact Array (ATCA) since 2004. The most densely time-sampled datasets are at $C$ and $X$ bands observed simultaneously (centred at $5.5$ and $9.0$~GHz, respectively). We downloaded the whole $X$-band dataset from the ATCA Calibrator Database\endnote{\url{https://www.narrabri.atnf.csiro.au/calibrators/calibrator_database_viewcal?source=0346-279}, accessed on 28 May 2026} for flux density light \mbox{curve modelling.}

PKS~0346$-$27 was also frequently observed with VLBI using the U.S. Very Long Baseline Array (VLBA) at multiple bands from June 1996 to October 2022. These observations were usually performed in the framework of calibrator and other surveys \cite{2002ApJS..141...13B,2009AJ....137.3718L, 2012AA...544A..34P,2016AJ....151..154G,2019MNRAS.485...88P,2021AJ....161...14P,2021AJ....162..121H,2025ApJS..276...38P}, and most of the data were recorded in $S$ and $X$ bands (around $2$ and $8$~GHz frequencies, respectively). The VLBI images and calibrated visibility data are publicly available from the Astrogeo\endnote{\url{https://astrogeo.org/cgi-bin/imdb_get_source.csh?source=J0348-2749}, accessed on 28 May 2026} database \cite{2025ApJS..276...38P}. 

In a fortunate coincidence, one of the $X$-band VLBA observations of PKS~0346--27 was conducted on 5 November 2021 (VLBA project code: UH007o), just two days after the large $\gamma$-ray flare detected by H.E.S.S. \cite{2026A&A...706A.246H}. We downloaded the calibrated and self-calibrated $X$-band visibility data of this epoch from the Astrogeo database, in order to make an image of the source. The short snapshot observation with an on-source time of $1$~s was performed in right circular polarization with the total bandwidth of $384$~MHz, divided into $12$ intermediate frequency channels, each $32$ MHz wide. From the ten-element VLBA interferometer, only the Mauna Kea radio telescope was missing. 

The visibility data were imaged in the Difmap \cite{1994BAAS...26..987S} software (version 2.5q). We performed standard hybrid mapping, which includes iterations of the \texttt{CLEAN} deconvolution algorithm \cite{1974A&AS...15..417H} and phase-only self-calibration \cite{1984ARA&A..22...97P}, followed by a few rounds of amplitude and phase self-calibration. Repeating the self-calibration on the already self-calibrated data cannot cause any bias in the result but may slightly improve the quality of the final image. Finally, circular Gaussian brightness distribution model components were fitted to the self-calibrated visibility data \cite{1995ASPC...82..267P}, to quantitatively characterize sizes and flux densities of the core and \mbox{jet components.} 

The long-term $X$- and $U$-band VLBA data series available in the Astrogeo archive are suitable for measuring the possible apparent jet proper motion in PKS~0346$-$27. The $X$-band ($7.3-8.4$~GHz) data coverage starts in August 1997 and lasts until October 2022. There is a longer gap in the $X$-band time sampling from 2019 to 2021. However, this interval is covered with $U$-band ($15$~GHz) VLBI measurements. We also downloaded all these $X$- and $U$-band datasets to image the source and model its brightness distribution on milliarcsecond (mas) angular scales, overall in $46$ individual epochs. The large volume of data was processed using an automatic Difmap imaging script\endnote{\url{https://github.com/rstofi/VLBI_Imaging_Script}, accessed on 28 May 2026} \cite{2017IAUS..324..247R}, to quickly obtain images at each epoch. The script follows a similar hybrid mapping procedure as described above. In a few individual epochs, when we found it necessary, we repeated the imaging manually in Difmap for better fidelity. In most of the epochs, a single jet component within $\sim$$2$~mas separation from the core component was found. Occasionally, when the array configuration and the higher sensitivity allowed us imaging more diffuse and extended emission, fainter outer jet components also appeared.

\subsection{Modelfit Uncertainties}

The uncertainties of the fitted model component parameters were estimated following a statistical approach \cite{1999ASPC..180..301F}. According to the standard practice \cite{2011A&A...526A..74M,2012AA...544A..34P,2018MNRAS.473.1388M,2021ApJ...919...40P}, $10\%$ absolute VLBI amplitude calibration uncertainty was added in quadrature to the formal error of the fitted flux density of each component. For snapshot observations, typically with short ($\sim$$1$~min) observing time and sparse $(u,v)$ coverage, the uncertainty of the jet component separations can be conservatively estimated as $20\%$ of the full width at half-maximum (FWHM) size of the elliptical Gaussian synthesised beam measured along the component position angle \cite{2009AJ....138.1874L,2013AJ....146..120L,2021ApJ...919...40P}. Table~\ref{tab:results} contains the log of the VLBI observations analysed, and the modelfit results for the component flux densities, sizes, and core--jet component separations, along with their calculated uncertainties.

\section{Results}
\subsection{Doppler Factor Estimate from the VLBI Measurement}  
\label{doppler_vlbi}

\begin{figure}[H]
\includegraphics[width=10.0 cm]{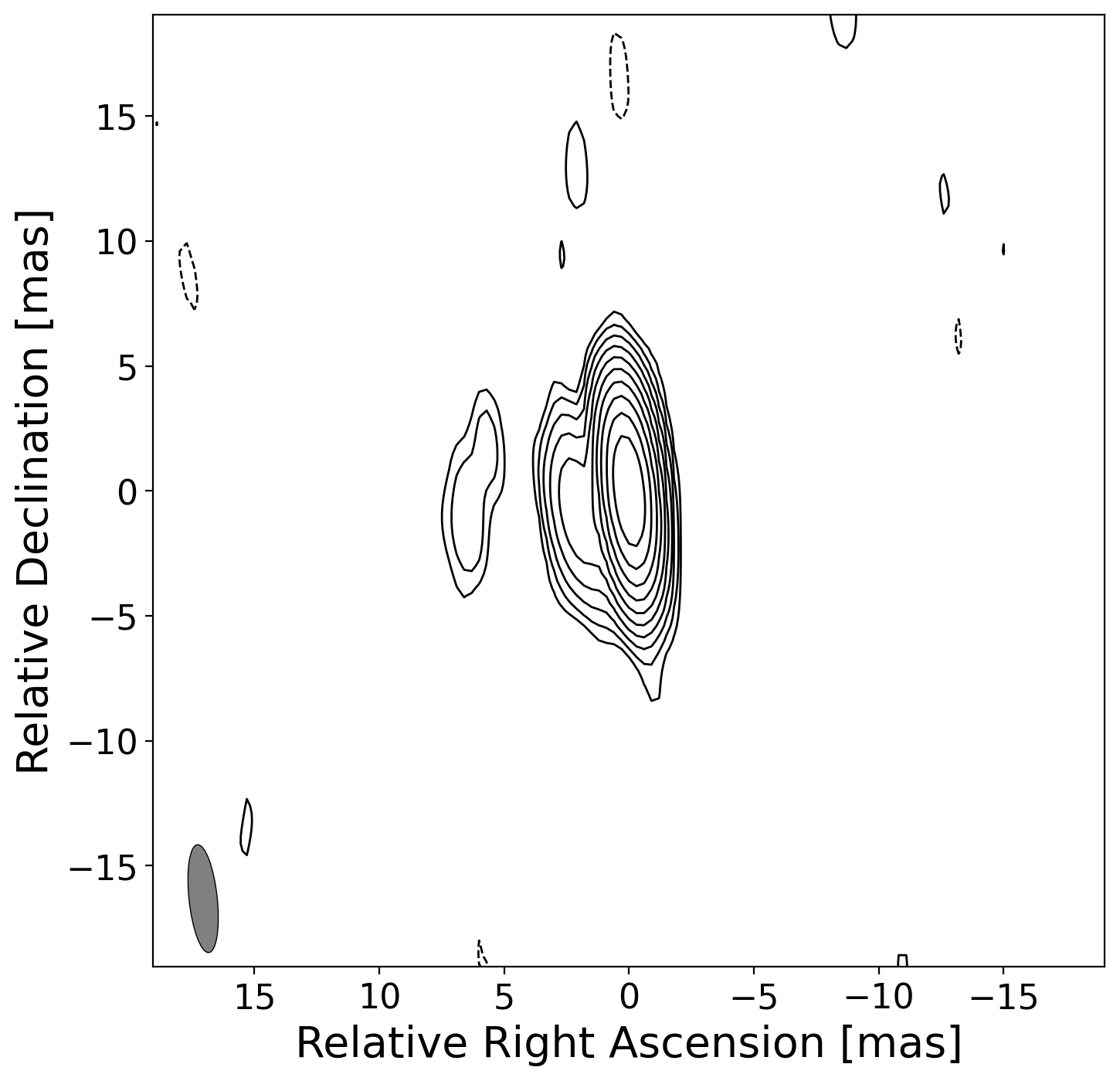}
\caption{The naturally weighted 8.7 GHz image of PKS~0346$-$27 from the VLBA experiment UH007o on 5 November 2021. The peak intensity is $2.12$~Jy\,beam$^{-1}$, with the lowest contours drawn at $\pm5.3$~mJy\,beam$^{-1}$. The positive  contour levels increase by a factor of 2. The size of the restoring beam is $4.3\,\mathrm{mas} \times 1.1\,\mathrm{mas}$ (FWHM) at the position angle $\mathrm{PA}=6.1^{\circ}$ measured from north through east, as indicated in the bottom-left corner.}
\label{vlbiimage}
\end{figure}

The quasi-simultaneous VLBA measurement gives an excellent opportunity to estimate the Doppler factor during the TeV flaring event. The naturally weighted 8.7 GHz VLBA image is shown in Figure~\ref{vlbiimage}. The brightness distribution is modelled with two circular Gaussian components, a core and a jet. The latter is located $(2.26 \pm 0.90)$~mas {east} from the core. The fitted core flux density is $(2.23 \pm 0.27)$~Jy, and the FWHM size $(0.32 \pm 0.02)$~mas. 

\textls[-25]{The core flux density and size can be used to estimate the core brightness temperature \cite{1982ApJ...252..102C}:}
\begin{linenomath}
\begin{equation}  \label{vlbiTb}
T_\mathrm{b, VLBI} = 1.22 \times 10^{12} (1 + z) \frac{S}{\phi^2\nu^2},
\end{equation}
\end{linenomath}
where $S$ is the flux density measured in Jy, $\phi$ the FWHM of the fitted circular Gaussian model component in mas, and $\nu$ the observing frequency in GHz. The core $T_\mathrm{b, VLBI}$ of PKS~0346$-$27 on 5 November 2021 was $(6.9 \pm 1.1) \times 10^{11}$~K. The component is considered to be resolved with the VLBA as the fitted FWHM exceeds the minimum resolvable angular size of the interferometer, according to Equation~(2) of \cite{2005AJ....130.2473K}.

The Doppler factor based on the VLBI core measurements can be calculated as  
\begin{linenomath} 
\begin{equation}\label{vlbidelta}
\delta_\mathrm{VLBI} = \frac{T_\mathrm{b,VLBI}}{T_\mathrm{b,int}},
\end{equation}
\end{linenomath}
where $T_\mathrm{b,int}$ denotes the intrinsic brightness temperature.
Following \cite{1994ApJ...426...51R}, this value is usually assumed to be equal to the equipartition brightness temperature, $5 \times 10^{10}$~K. It should be noted that $T_\mathrm{b,int}$ can be variable in time, especially in a flaring source like ours. Studies of jetted AGN samples \cite{2006ApJ...642L.115H,2021ApJ...923...67H} showed that $T_\mathrm{b,int}$ can generally be somewhat lower than the equipartition value ($\sim$3--4 $\times$ $10^{10}$~K) in most of the sources. Our observed 8.7~GHz frequency corresponds to $\sim$17~GHz rest-frame frequency at the source redshift, and because $T_\mathrm{b,int}$ is known to be frequency-dependent \cite{2014JKAS...47..303L,2020ApJS..247...57C}, it could be around $3 \times 10^{10}$~K. \mbox{However, Ref. \cite{2006ApJ...642L.115H,2021ApJ...923...67H}} also found that  $T_\mathrm{b,int}$ can reach $\sim$$10^{11}$~K in a flaring state when energy equipartition conditions between particles and magnetic field are not met. Given that PKS~0346$-$27 was in outburst in November 2021, we adopt $T_\mathrm{b,int} = 10^{11}$~K to calculate the Doppler factor according to Equation~\eqref{vlbidelta}. This way, we arrive at a conservative estimate of the Doppler factor, \textbf{$\delta_\mathrm{VLBI} = 7.0$}.

\begin{figure}[H]
        \includegraphics[width=0.8\linewidth]{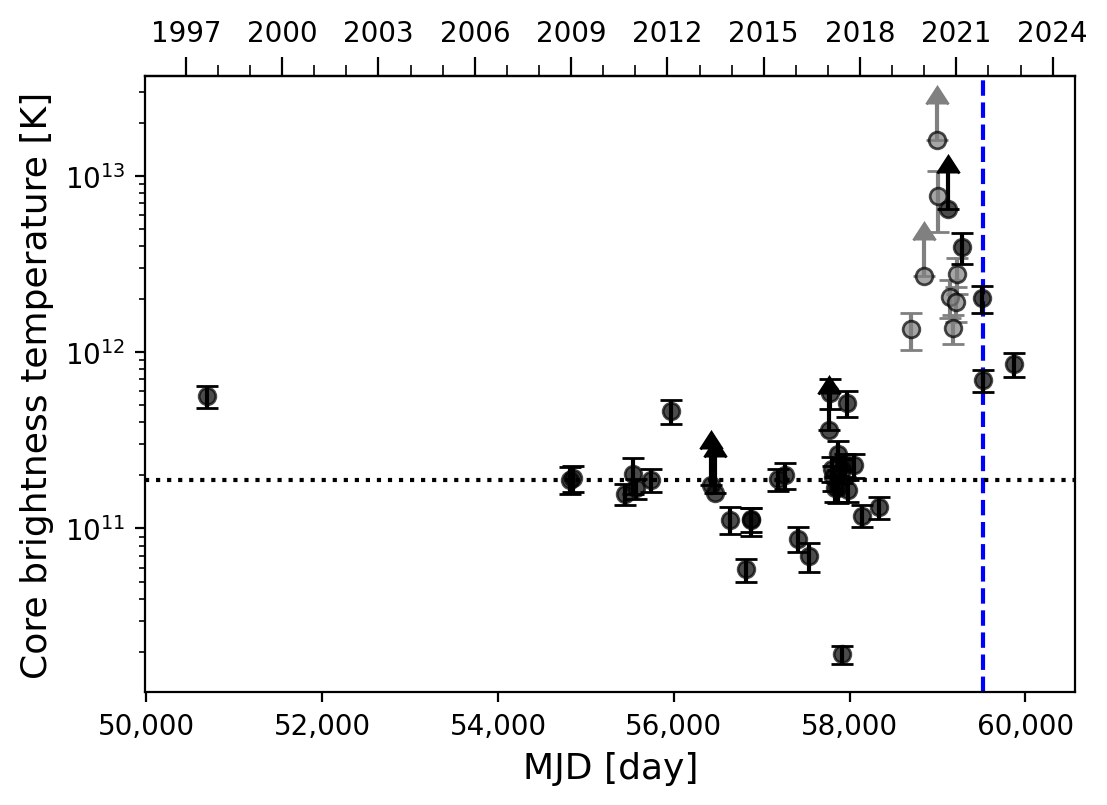}
    \caption{The brightness temperature of the core of PKS~0346$-$27 as a function of time. The lower limit of $T_\mathrm{b}$ is plotted with an upward arrow if the source was unresolved with the interferometer. The dotted horizontal line indicates the median brightness temperature during quiescence, $1.9 \times 10^{11}$~K, calculated from the $X$-band data (black symbols). The grey-coloured symbols represent the $U$-band archival VLBA data. The blue vertical dashed line marks the date of the TeV flaring event.}
    \label{fig:tb}
\end{figure}

The Doppler factor during the quiescent (non-flaring) state can also be estimated using the whole series of $X$-band VLBA data. We calculated the core brightness temperature at all epochs available (Table~\ref{tab:results}). Lower limit to $T_\mathrm{b}$ was considered when the source was unresolved by the interferometer. Figure~\ref{fig:tb} shows the core $T_\mathrm{b}$ as a function of time. According to the plot, the quiescent state lasted until the beginning of 2019 with median core brightness temperature $T_\mathrm{b,median} = 1.9 \times 10^{11}$~K during this period. As mentioned above, at $\sim$17~GHz, $T_\mathrm{b,int} \approx 3 \times 10^{10}$~K can be considered in quiescence. With this assumption, the Doppler factor during the quiescent state is $\delta_\mathrm{VLBI,\,q} \approx 6$. {The Doppler factors are usually higher in flaring state than in quiescence (e.g., \cite{2025A&ARv..33....8R}). Our findings are consistent with this.}

\subsection{Doppler Factor Estimate from Radio Variability} 
\label{doppler_vari}

\begin{figure}[H]
\includegraphics[width=12.0 cm]{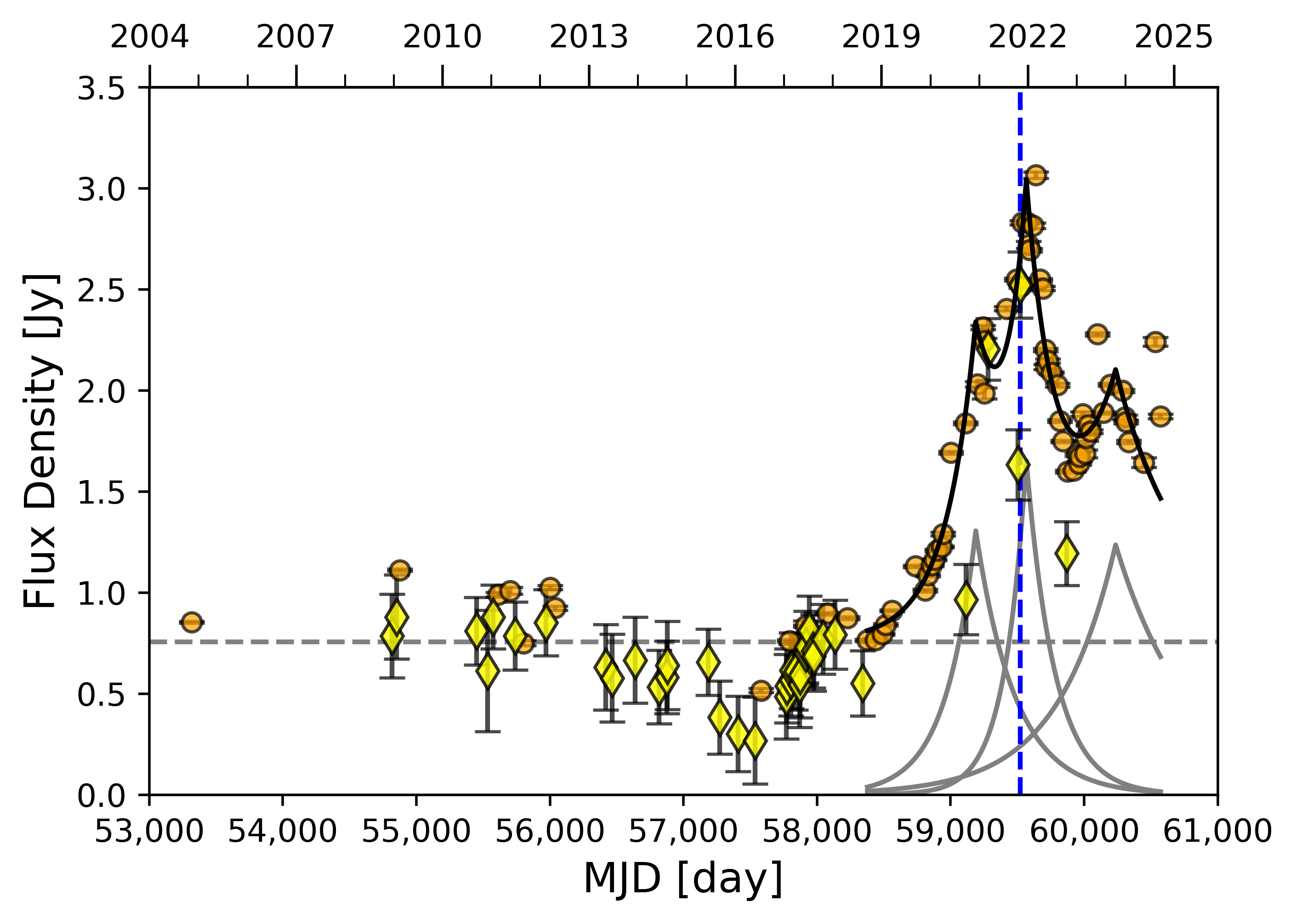}
\caption{The ATCA (orange circles) and the VLBA (yellow diamonds) $X$-band radio light curve of PKS~0346$-$27 from November 2004 to September 2024. The black curve is the fitted 3-component flare model and the grey curves represent the individual components. The grey horizontal dashed line shows the median flux density in the quiescent state, $0.76$\,Jy. The blue vertical dashed line marks the date of the TeV flaring event.}
\label{fig:lightcurve}
\end{figure}

The Doppler factor can also be estimated through radio variability by fitting the total flux density light curve of flares \cite{1999ApJS..120...95V,2009A&A...494..527H}. The fitted exponential curve is described as
\begin{linenomath}
\begin{equation}  \label{flarefunction}
\Delta S(t) =
    \begin{cases}
        \Delta S_\mathrm{max}e^{(t-t_\mathrm{max})/\tau} \,\mathrm{if}\, t < t_\mathrm{max} \\
        \Delta S_\mathrm{max}e^{(t_\mathrm{max}-t)/1.3\tau} \,\mathrm{if}\, t > t_\mathrm{max},
    \end{cases}
\end{equation}
\end{linenomath}
where $t$ is the time, $t_\mathrm{max}$ is the epoch of the maximum flux density, $\Delta S_\mathrm{max}$ is the maximum amplitude of the flare in Jy, and $\tau$ is the rise time of the flare in days\endnote{We caution that Equation~(1) in \cite{2009A&A...494..527H} is mistyped, not matching the equation originally presented in \cite{1999ApJS..120...95V}.}. The $X$-band VLBA data coverage happens to fill the gap in the 9 GHz ATCA monitoring data during the years 2013--2016 (Figure~\ref{fig:lightcurve}). In most of the overlapping periods, the VLBA flux density measurements (core and jet components combined) provide values somewhat below but close to the total flux density measured with ATCA. It indicates that most of the radio emission originates from the compact structure revealed with VLBI. 
Concerning flux density variability, it is feasible to treat the two $X$-band datasets together. Before fitting the light curve for flares according to Equation~\eqref{flarefunction}, we subtracted a base value from the flux densities. It was determined as the median of the flux densities measured during the quiescent state, $0.76$~Jy. The fit provided the $\Delta S_\mathrm{max}$ and $\tau$ parameters.

{{We made attempts to fit the flaring section of the light curve with one, two, and three flare component models. (Fits with with more than three flare components were not successful.) Out of the three variants, the three-component model gave the lowest reduced $\chi^2$ value ($2.7$) and the smoothest residuals. We chose this three-component model because it is the best to reproduce the shape of the light curve.}} 

The fitted three distinct flare components are similar in magnitude (see the grey curves in Figure~\ref{fig:lightcurve}, and their sum in black). The closest radio flare in time to the TeV flare is the middle one with $t_\mathrm{max}$ on 18 December {{2021}}. We consider the parameters of this flare component when calculating the variability Doppler factor, $\delta_\mathrm{var}$. Here $\Delta S_\mathrm{max} = (1.65 \pm 0.01)$~Jy and $\tau = (164 \pm 2)~{\mathrm{d}}$.

The variability brightness temperature can be calculated as
\begin{linenomath}
\begin{equation}  \label{varTb}
T_\mathrm{b,var} = 1.548 \times 10^{-32} \frac{\Delta S_\mathrm{max} D^2_\mathrm{L}}{\nu^2\tau^2(1+z)},
\end{equation}
\end{linenomath}
where $D_\mathrm{L}$ is the luminosity distance of the source in m, and $\nu$ is the observing frequency in GHz \cite{2000A&AS..141..221Q,2009A&A...494..527H,2025A&ARv..33....8R}. In this equation, {$\Delta S_\mathrm{max}$ is given in Jy and $\tau$ in days.}
We obtain {$T_\mathrm{b,var} = (2.4 \pm 0.2) \times 10^{14}$~K}. The variability Doppler factor is calculated as 
\begin{linenomath} 
\begin{equation}\label{vardelta}
\delta_\mathrm{var} = \left( \frac{T_\mathrm{b,var}}{T_\mathrm{b,int}} \right) ^{\frac{1}{3}},
\end{equation}
\end{linenomath}
assuming the same intrinsic brightness temperature as for estimating $\delta_\mathrm{VLBI}$, $T_\mathrm{b,int} = 10^{11}$~K (Section~\ref{doppler_vlbi}). Thus, the variability Doppler factor is {$\delta_\mathrm{var} = 13.4$, more than twice the quiescent value $\delta_\mathrm{VLBI,\,q}$}.

\subsection{Proper Motion in the Jet} \label{propmotion}
The long-term $X$-band VLBI data coverage allowed us to identify three different jet components east of the core during the $25$ yr period investigated. The component closest to the core (J1) has about 2--2.5~mas separation (Table~\ref{tab:results}). The two outer components, J2 and J3, are located at $\sim$3.5 and $\sim$4.5~mas from the core, respectively. These two components are fainter and more diffuse, and they could be identified at a few epochs only. Moreover, the component identification is uncertain in one case, at the earliest epoch of 27 August 1997, which is detached from the bulk of the data starting to accumulate from 2008. The single component found in 1997 can either be J1 or J2. We treat it as an early appearance of J2 that could be securely detected two decades later, after moving further away from the core. Certainly, more VLBI imaging data between 1997 and 2008 would have helped the component identification.

\begin{figure}[H]
        \includegraphics[width=0.8\linewidth]{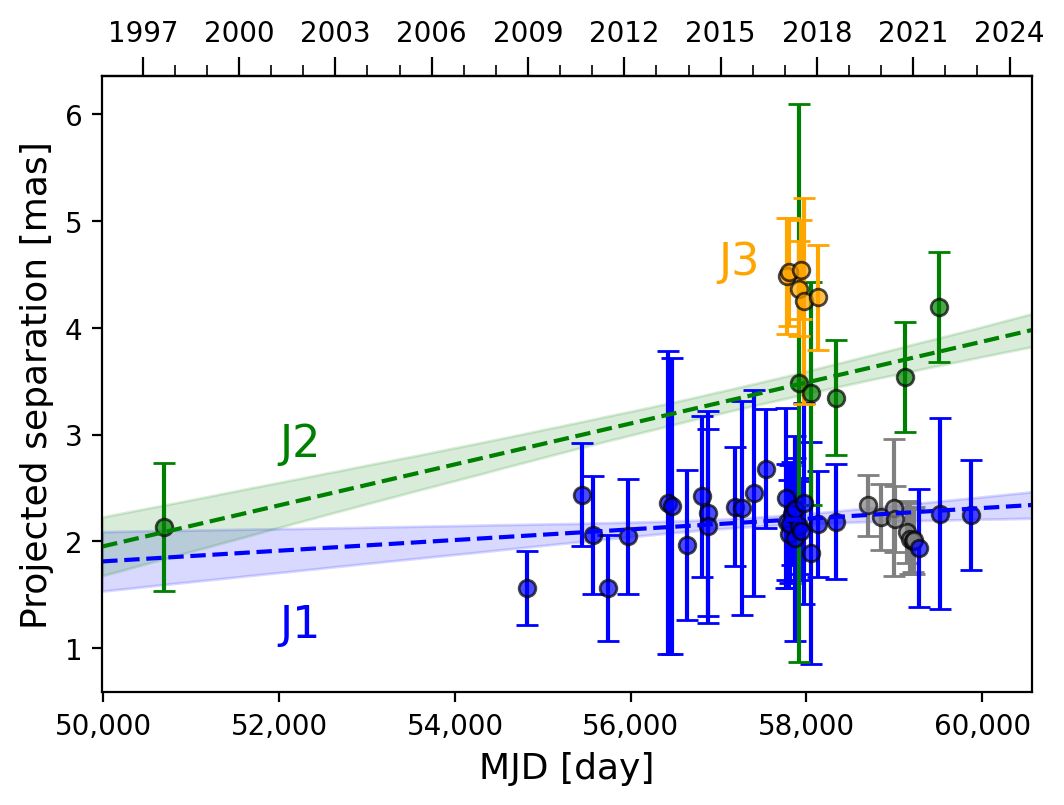}
    \includegraphics[width=0.8\linewidth]{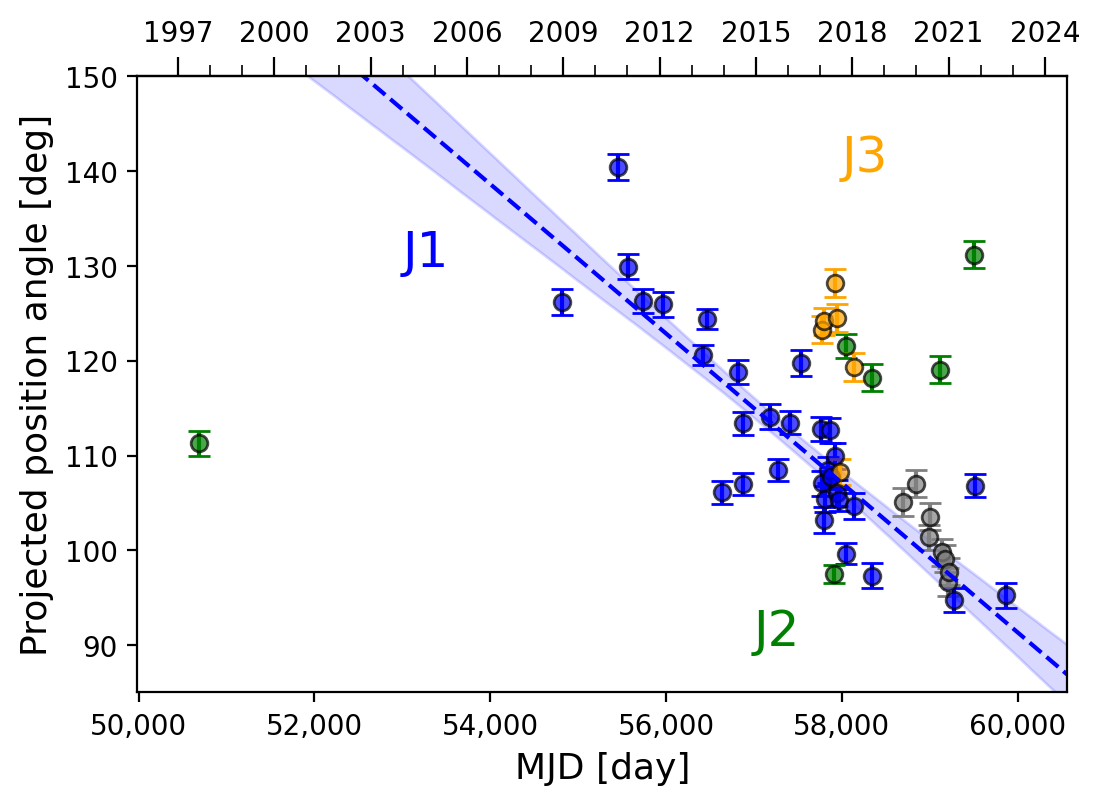}
    \caption{Proper motion plots of PKS~0346$-$27 jet components based on the archival $X$-band VLBA data. The dashed lines show the best-fit linear regression. The shaded areas represent the $1\sigma$ uncertainty of each fit. The colouring of the jet components identified is the following: J1---blue, \mbox{J2---green}, J3---orange. The grey-coloured points represent the $U$-band archival VLBA data for component J1. These are shown for illustration purposes only and not used in the fits. \textit{Top:} Core--jet separation and radial proper motion of the jet components. \textit{Bottom:} Jet component position angles with respect to the core, measured from north through east as a function of time.}
    \label{fig:propermotion}
\end{figure}

The apparent proper motions of the identified jet components were calculated based on the relative positions of the fitted circular Gaussian brightness distribution models with respect to the core. The angular separation vs. time ($\mu_{\rm r}$, i.e., the apparent {radial} proper motion) and the position angle vs. time ($\mu_{\rm PA}$) were fitted with linear functions. For the separation, we fitted the J1 and J2 components, while for the position angles, only the J1 component was fitted. In the other cases, reliable fits were not feasible due to the scarce time sampling or the large scatter of the data points. Figure~\ref{fig:propermotion} shows the core--jet component separations and the position angles (PAs) as a function of time, along with the best-fit linear regression. The PA of the jet component seen at the earliest epoch of 27 August 1997 clearly falls outside the trend observed for J1 and is more consistent with the values found later for J2. This supports the assumption that this component is an early appearance of J2 and not J1.

By imaging the $U$-band (15 GHz) data, we successfully identified the J1 component at $\sim$$2$~mas from the core in all eight available epochs. These positions are also plotted in Figure~\ref{fig:propermotion}, but for illustration purposes only. They were not used for the proper motion determination. The outer components were completely resolved out at this higher frequency. 

\textls[-15]{The apparent {speed} of a jet component in units of the speed of light $c$ can be calculated as}
\begin{equation} \label{eq:beta}
    \beta_\mathrm{app} = 0.0158 \, \frac{\mu D_{\rm L}}{1+z}, 
\end{equation}
where {$\mu$ is the component angular proper motion} given in $\mathrm{mas\,yr}^{-1}$ and $D_\mathrm{L}$ in Mpc units \cite{2012ApJ...760...77A}. The $\mu_{\rm r,\,J1}$ for component J1 is $(0.018 \pm 0.013)~\mathrm{mas\,yr}^{-1}$, which corresponds to $\beta_\mathrm{app,\,r,\,J1} = 0.94 \pm 0.64$. The rate of the change in the J1 position angle is $\mu_{\rm PA,\,J1} = (-2.87 \pm 0.33)^{\circ}\,\mathrm{yr}^{-1}$, resulting in a total of $\sim$$65^{\circ}$ change over the entire 25 yr time span. In other words, the direction of the innermost jet changed from southeastern to nearly eastern. {Because of this significant annual change in the position angle, the transverse velocity component must also be taken into account in addition to the radial component. The resultant total proper motion ($\mu_\mathrm{t}$) was calculated as the angular distance between the positions of the J1 component at the first and last observed epochs ($\sim$$1.2$~mas) divided by the time that passed between them ($13.8$~yr). The proper motion is thus $\mu_\mathrm{t} = (0.088 \pm 0.015)~\mathrm{mas\,yr}^{-1}$, which corresponds to $\beta_\mathrm{app,\,J1} = 4.55 \pm 0.78$ apparent speed.}
The radial proper motion of component J2, $\mu_{\rm r,\,J2} = (0.069 \pm 0.012)~\mathrm{mas\,yr}^{-1}$, appears higher than that of J1. {However, this value is somewhat lower than, but consistent within the errors with the total proper motion $\mu_\mathrm{t}$ derived for J1. The radial proper motion of J2 translates} to $\beta_\mathrm{app,\,r,\,J2} = 3.60 \pm 0.62$. There is not any discernible trend in the position angles of J2 and J3.

\subsection{Calculating the Bulk Lorentz Factor} \label{lorentz}

Knowing the apparent {speed} and the Doppler factor, we can calculate the bulk Lorentz factor, the physical parameter that characterizes the plasma flow in the jet. Following \cite{1993ApJ...407...65G}, the bulk Lorentz factor of a jet is given as 
\begin{equation} \label{eq:lorentz}
    \Gamma = \frac{\beta_\mathrm{app}^2 + \delta^2 + 1}{2\delta}.
\end{equation} 

We can calculate $\Gamma_\mathrm{VLBI}$ by substituting the VLBI Doppler factor $\delta_\mathrm{VLBI}$, and $\Gamma_\mathrm{var}$ by using the variability Doppler factor $\delta_\mathrm{var}$ for $\delta$ in Equation~\eqref{eq:lorentz}. 
The detection of J1 is the most robust, and this component is the closest to the synchrotron-self-absorbed base of the jet (i.e., the core), {therefore, we select its resultant apparent speed, $\sim$$4.6\,c$, for calculating $\Gamma_\mathrm{VLBI}$ and $\Gamma_\mathrm{var}$ (Table~\ref{tab:parameters}). However, substituting either the apparent radial speed of J1, $\sim$$1\,c$, or the apparent radial speed of J2, $\sim$$3.6\,c$, makes little difference in the calculated $\Gamma$.}

\begin{table}[H]
        \caption{Calculated physical parameters of the PKS~0346$-$27 jet from VLBI imaging and radio variability.}
    \label{tab:parameters}
\begin{tabularx}{\textwidth}{CCCC}
    \toprule
        \textbf{Method} &  \textbf{\textit{T}\textsubscript{b} [K]}            & \boldmath{$\delta$}       & \boldmath{$\Gamma$} \\
        \midrule
        VLBI   & $(6.9 \pm 1.1) \times 10^{11}$ & $7.0$ & $5.0$ \\
        variability   & $(2.4 \pm 0.2) \times 10^{14}$ & $13.4$ & $7.5$   \\
        \bottomrule
    \end{tabularx} 

\noindent{\footnotesize{Notes:  {{Col.~1}}---estimation method; {Col.~2}---brightness temperature; {Col.~3}---Doppler factor; {Col.~4}---Lorentz factor.}}
\end{table}

\section{Discussion}
\label{discussion}

According to VLBI kinematic measurements, most of the AGN jets have an apparent speed of $2 < \beta_\mathrm{app} < 30$ \cite{1988ApJ...329....1C,1994ApJ...430..467V,1999NewAR..43..757K,2019ApJ...874...43L}. The apparent jet speeds we determined for PKS~0346$-$27 are relatively small compared to other sources. It has been shown that jets in TeV blazars tend to be considerably slower than in other blazars \cite{2018ApJ...853...68P}. Their measured apparent speeds are consistent with no motion, and some of them may show stationary features. PKS~0346$-$27 also fits in this picture. The moderate value measured for the innermost jet component, $\beta_\mathrm{app,\,r,\,J1} \approx 1$, could be caused by projection effects. This notion is supported by the high rate of change in the position angle {and the high resultant apparent speed, suggesting a significant bending in the jet. However, the current data are not sufficient to attempt the modelling of possible jet precession. We cannot rule out that J1 is a standing shock. In this case, it does not represent the jet kinematics, and the calculations should be based on $\beta_\mathrm{app,\,r,\,J2}$ rather than the apparent speed of the J1 component.}

The Doppler factor and the Lorentz factor in PKS~0346$-$27 were estimated using two different methods, based on radio variability and VLBI imaging. Table~\ref{tab:parameters} summarizes the calculated physical parameters both by the VLBI imaging and radio variability methods. Considering the formal errors in the measured parameters and the inherent uncertainty in assuming an actual value for $T_\mathrm{b,int}$ in Equations~\eqref{vlbidelta} and \eqref{vardelta}, as well as possible time variability, both estimates of the Lorentz factor are consistent with $\Gamma \lesssim 10$. {It should be noted that Equation \eqref{eq:lorentz} provides estimate under the assumption of a constant jet direction (radial jet motion). The bending of the jet therefore introduces an additional uncertainty in the parameters derived. The inferred Lorentz factors should be interpreted as average values. However, the overall trends remain unchanged, and the jet curvature does not affect the main conclusions of our analysis.}

The $\delta_\mathrm{VBLI}$ was calculated from VLBI observation taken almost simultaneously with the TeV flaring event in November 2021. This measurement allows us to derive the radio properties of the source at the same time as the $\gamma$-flare, which is a rare opportunity. The publicly available archival VLBA data spanning $\sim$$25$ years allowed us to also determine the core $T_\mathrm{b}$ in the quiescent state and therefore estimate the Doppler factor during quiescence. The value $\delta_\mathrm{VBLI,\,q} \approx 6$ is {slightly lower than} the one calculated during the $\gamma$-ray flare. The brightening in the radio starts in $2019$ and can be clearly seen in the $T_\mathrm{b}$ plot (Figure~\ref{fig:tb}), consistently with the behaviour of the total flux density light curve (Figure~\ref{fig:lightcurve}). {We note that the peak in the core $T_\mathrm{b}$ curve does not coincide with the flux density peak, with $\sim$$1$~yr difference between them. The brightness temperature (Equation~\eqref{vlbiTb}) depends not only on the flux density but the size of the emission region as well. At the time of the TeV event, $T_\mathrm{b}$ was lower than before (Figure~\ref{fig:tb}), which can be explained by the increased source size.}

Modelling the frequently-sampled combined (ATCA and VLBA) $X$-band radio light curve reveals three peaks within one year of the TeV flare. Flare modelling is sensitive to the input parameters, especially when there are multiple flare components close to each other in time. Our best-fit model characterised by the lowest $\chi^2$ value includes two almost identically-shaped flare components and a third flare that is brighter. The middle, brightest flare is the closest to the TeV event, following it by 46 days. In turn, the first modelled radio flare occurs 333 days before the TeV $\gamma$-flare, and the third one {$715$ days} after (Figure~\ref{fig:lightcurve}).

 We used the parameters of the middle modelled radio flare in our calculations, since it is the closest in time to the TeV event.  {Among the other modelled flare components, we find that the first one precedes, and therefore is unlikely to be associated with the subsequent TeV outburst. The third flare with its potential $\sim$2 yr delay could in principle be related to the TeV event. Given the frequent radio flaring activity in PKS~0346$-$27, it is possible that the coincidence of the TeV flare and the middle radio flare is accidental. The source showed strong $\gamma$-ray variability even before 2019 \cite{2018ATel11251....1A}, during a quiescent period in the radio. We cannot rule out the possibility that the radio flare is related to one of the previous $\gamma$-ray outbursts. The relatively short time difference between the TeV and $X$-band radio flares could indicate co-spatial radio and TeV $\gamma$-ray emission regions. Proving the co-spatiality would require synchrotron-self-Compton modelling in the different phases of source variability (e.g., \cite{2012A&A...539A.149H,2014ApJ...782...13A}).}

Radio flares usually show time delay with respect the $\gamma$-ray flares. This can amount to days or even years \cite{2018MNRAS.473.4107P}. Such a delay is interpreted as the time needed for the plasma to travel along the jet to the distance at which the radio emission becomes optically thin. A recent study of blazar jets \cite{2026arXiv260403847K} found that the typical time delay {between a $\gamma$-ray outburst in the 100 MeV$-$500 GeV range and a radio flare in the GHz radio regime is $\sim$$180$\,days. For a TeV-energy outburst, this delay may be shorter when it originates from the same shock region as the radio flare. However, the TeV--radio correlation is still poorly understood because there is no clear TeV--radio correlation found, unlike in the GeV bands \mbox{(e.g., \cite{2011ApJ...727..129A,2012A&A...539A.149H,2014ApJ...782...13A,2025A&ARv..33....8R}).}} 
In particular, in the case of the blazar S5 0716$+$714, the radio flare followed the high-energy {(GeV $\gamma$-ray)} outburst $\sim$$65$\,days later \cite{2013A&A...552A..11R}, similarly to our finding for PKS~0346$-$27. {By investigating PKS~0346$-$27 together with many other quasars, \cite{2026arXiv260403847K} found a $(588 \pm 20)$~d time lag between the GeV and GHz bands. However, that GeV emission peak happened in early 2019.}

From their comparisons using large AGN samples, it is known that Doppler-boosting factors determined from VLBI imaging and from radio flux density variability have similar average values on population levels but can significantly differ for individual objects \cite{1999ApJ...521..493L,2015MNRAS.454.1767L}, with the variability Doppler factors being the more reliable in general \cite{1999ApJ...521..493L}. Our estimates for $\delta_\mathrm{var}$ and $\delta_\mathrm{VBLI}$ in PKS~0346$-$27 agree {within a factor of $\sim$$2$}, which is not unprecedented in the literature (e.g., \cite{2010A&A...521A...6V}). The difference between the estimated Doppler factors can be related to the fact that the two measurements are not made at the same time and at the same frequency. Also uncertain is the value adopted for the intrinsic brightness temperature, which could also be variable in time \cite{2006ApJ...642L.115H,2021ApJ...923...67H}.

The recent simultaneous multi-wavelength study of PKS~0346$-$27 \cite{2026A&A...706A.246H} that prompted our radio analysis presented here resulted in the estimation of the bulk Lorentz factor. The SED of PKS~0346$-$27 was modelled using a single-zone, time-independent hadronic approach, which gave $\Gamma = 10$. On the other hand, the attempt for assuming a single-zone leptonic model yielded $\Gamma \gtrsim 80$ \cite{2026A&A...706A.246H}. However, the authors discarded the latter model due to other extreme parameter values obtained. The $\Gamma \lesssim 10$ values we independently estimated from radio data are consistent with and support the hadronic model proposed in \cite{2026A&A...706A.246H}. 
For samples of jetted AGN observed with VLBI, the distribution of bulk Lorentz factors peaks between $5-15$ \cite{2019ApJ...874...43L}. In some cases, the values can reach $\Gamma \approx 40$, but $\Gamma \gtrsim 80$ would clearly be too high.

\section{Summary and Conclusions}
\label{conclusions}

PKS~0346$-$27 is the second most distant source ($z=0.991$) known to show a VHE $\gamma$-ray outburst. Its TeV-energy flare occurred on 3 November 2021. By modelling the broadband SED observed at the time of this flare, \cite{2026A&A...706A.246H} proposed a single-zone hadronic model to describe the physical parameters of the jet. An alternative single-zone leptonic model could also provide a good fit to the SED but was discarded on the basis of the extremely high Doppler and bulk Lorentz factors \cite{2026A&A...706A.246H}.

Here we estimated the jet Doppler factor by two other independent methods from archival radio data. One is based on a VLBI observation very close in time to the TeV flare, and another is on modelling long-term radio flux density monitoring data. These Doppler-factor estimates were then applied to determine the bulk Lorentz factor. The high-resolution radio image, derived from the 8.7 GHz VLBA measurement on 5 November 2021, shows a bright ($\sim$2~Jy) radio core and a jet component located at $\sim$2~mas towards the east. By fitting Gaussian brightness distribution model to the core, assuming $10^{11}$~K as the intrinsic brightness temperature characteristic to jets during outbursts, and using the apparent jet speed measured with VLBI at $8$~GHz on a decade-long time scale, we arrived at a \textbf{$\Gamma \approx 5$} Lorentz-factor estimate.  

We also analysed the $X$-band radio light curve. According to our flare modelling, three major brightening events have occurred since 2020, the second being the closest in time to the $\gamma$-ray flaring event. From the variability Doppler factor, we estimated \textbf{$\Gamma=7.5$}. 

The results of both radio-based Lorentz-factor estimates support the single-zone hadronic model proposed in \cite{2026A&A...706A.246H}, where $\Gamma=10$ was estimated from SED modelling, rather than the leptonic model with $\Gamma \gtrsim 80$. The bulk Lorentz factors derived for PKS~0346$-$27 from the radio data are comparable to the typical values found for other blazars. 

By presenting PKS~0346$-$27 as an example, our results demonstrate that analysing high-resolution radio imaging and/or radio flux density monitoring data can be a valuable addition to high-energy observations and SED modelling when determining the physical parameters of the jet. This case also highlights the importance of long-term VLBI and radio total flux density monitoring of prominent blazars. Such archival data could be invaluable at later times when an extraordinary event, a $\gamma$-ray flare (like in or case) or, e.g., a neutrino detection is associated with any particular source.  

\vspace{6pt}

\authorcontributions{Conceptualization, S.F.; methodology, M.K., S.F. and K.\'E.G.; formal analysis, M.K.; writing---original draft preparation, M.K.; writing---review and editing, S.F. and K.\'E.G.; visualization, M.K.; supervision, S.F. All authors have read and agreed to the published version of the~manuscript.}
\funding{This research received no external funding.}
\dataavailability{The calibrated VLBI data underlying this article are available in the Astrogeo Center database at  
\url{https://astrogeo.org/cgi-bin/imdb_get_source.csh?source=J0348-2749}, accessed on 28 May 2026. The flux density monitoring data can be obtained from the ATCA Calibrator Database at \url{https://www.narrabri.atnf.csiro.au/calibrators/calibrator_database_viewcal?source=0346-279&detailed=true}, accessed on 28 May 2026.} 

\acknowledgments{{We thank the two anonymous reviewers for their insightful comments and suggestions that led to the improvement of the paper. M.K. also thanks Emma Kun for valuable advice.} The National Radio Astronomy Observatory is a facility of the National Science Foundation operated under cooperative agreement by Associated Universities, Inc. 
The Australia Telescope Compact Array is part of the Australia Telescope National Facility (\url{https://ror.org/05qajvd42}, accessed on 28 May 2026), which is funded by the Australian Government for operation as a National Facility managed by CSIRO. 
The authors acknowledge the use of Astrogeo Center database maintained by L. Petrov (\url{https://doi.org/10.25966/kyy8-yp57}, accessed on 28 May 2026). 
This work was supported by the HUN-REN Hungarian Research Network.}

\conflictsofinterest{The authors declare no conflicts of interest.} 

\abbreviations{Abbreviations}{
The following abbreviations are used in this manuscript:
\\

\noindent 
\begin{tabular}{@{}ll}
AGN & active galactic nuclei\\
ATCA & Australia Telescope Compact Array\\
FSRQ & flat-spectrum radio quasar\\
FWHM & full width at half-maximum \\
H.E.S.S. & High Energy Stereoscopic System\\
IC & inverse Compton\\
ICRF & International Celestial Reference Frame\\
IR & infrared\\
LAT & Large Area Telescope\\
mas & milliarcsecond \\
MJD & Modified Julian Date\\
PA & position angle\\
\textit{ROSAT} & \textit{Roentgen Satellite}\\
SED & spectral energy distribution \\
VHE & very high energy\\
VLBI & very long baseline interferometry \\
VLBA & very long baseline array \\
\end{tabular}
}
  

\startlandscape
\appendixtitles{yes} 
\appendix

\begin{adjustwidth}{+\extralength}{0cm}
\section{Data Table}
\end{adjustwidth}
\addtocounter{table}{-1}
\begin{table}[H]

    \caption{Details of the archival VLBI observing data obtained from the Astrogeo database and the modelfit results for the jet components. }
    
    \small
\setlength{\cellWidtha}{\textwidth/13-2\tabcolsep+0.2in}
\setlength{\cellWidthb}{\textwidth/13-2\tabcolsep-0in}
\setlength{\cellWidthc}{\textwidth/13-2\tabcolsep-0in}
\setlength{\cellWidthd}{\textwidth/13-2\tabcolsep-0in}
\setlength{\cellWidthe}{\textwidth/13-2\tabcolsep-0in}
\setlength{\cellWidthf}{\textwidth/13-2\tabcolsep-0.2in}
\setlength{\cellWidthg}{\textwidth/13-2\tabcolsep-0in}
\setlength{\cellWidthh}{\textwidth/13-2\tabcolsep-0.2in}
\setlength{\cellWidthi}{\textwidth/13-2\tabcolsep+0.2in}
\setlength{\cellWidthj}{\textwidth/13-2\tabcolsep-0in}
\setlength{\cellWidthk}{\textwidth/13-2\tabcolsep-0in}
\setlength{\cellWidthl}{\textwidth/13-2\tabcolsep-0in}
\setlength{\cellWidthm}{\textwidth/13-2\tabcolsep-0in}

\begin{tabularx}{\textwidth}{
>{\centering\arraybackslash}m{\cellWidtha}
>{\centering\arraybackslash}m{\cellWidthb}
>{\centering\arraybackslash}m{\cellWidthc}
>{\centering\arraybackslash}m{\cellWidthd}
>{\centering\arraybackslash}m{\cellWidthe}
>{\centering\arraybackslash}m{\cellWidthf}
>{\centering\arraybackslash}m{\cellWidthg}
>{\centering\arraybackslash}m{\cellWidthh}
>{\centering\arraybackslash}m{\cellWidthi}
>{\centering\arraybackslash}m{\cellWidthj}
>{\centering\arraybackslash}m{\cellWidthk}
>{\centering\arraybackslash}m{\cellWidthl}
>{\centering\arraybackslash}m{\cellWidthm}
}

    \toprule
\multicolumn{2}{c}{\textbf{Observing Date}} 
& \textbf{Project} 
& \textbf{Frequency} 
& \textbf{On-Source} 
& \textbf{Bandwidth} 
& \multirow{2}{*}{\textbf{Ref.}} 
& \multirow{2}{*}{\textbf{Comp.}} 
& \textit{\textbf{S}} 
& \boldmath{$\phi$} 
& \textbf{\textit{T}\textsubscript{b}}
& \textit{\textbf{R}} 
& {\textbf{PA}} \\ 

\textbf{yyyy-mm-dd}
& \textbf{[MJD]} 
& \textbf{Code} 
& \textbf{[GHz] }
& \textbf{time} \textbf{[min]} 
& \textbf{[MHz]} &  &  
& \textbf{[Jy]} 
& \textbf{[mas]} 
&  \textbf{[10\textsuperscript{11}} \textbf{K]}
& \textbf{[mas]}
 & \textbf{[}\boldmath{$^{\circ}$}\textbf{]} \\
\midrule
1997-08-27 & 50687
 & bb023 & 8.3 & 5 & 32 & \citep{2002ApJS..141...13B} & C & 1.337 (0.170) & 0.29 (0.02) & 5.61 (0.81) & &  \\
 &  &  &  &  &  &  & J2 & 0.120 (0.070) & 4.23 (0.23) & & 2.14 (0.60) & $-$68.7 (1.3) \\
2008-12-17 & 54817 & rdv72 & 8.6 & 52 & 64 &  & C & 0.629 (0.101) & 0.33 (0.03) & 1.90 (0.33) & &  \\
 &  &  &  &  &  &  & J1 & 0.085 (0.030) & 1.48 (0.11) & & 1.56 (0.34) & $-$53.8 (1.4) \\
2009-01-21 & 54852 & rdv73 & 8.6 & 3.8 & 64 &  & C & 0.800 (0.123) & 0.37 (0.03) & 1.93 (0.32) & &  \\
2010-09-11 & 55450 & bc191k(3) & 8.6 & 5.5 & 128 &  & C & 0.688 (0.088) & 0.38 (0.02) & 1.57 (0.22) & &  \\
 &  &  &  &  &  &  & J1 & 0.047 (0.017) & 0.89 (0.05) & & 2.44 (0.49) & $-$39.6 (1.4) \\
2010-12-05 & 55535 & s3111a & 8.6 & 1 & 128 &  & C & 0.556 (0.123) & 0.30 (0.04) & 2.04 (0.47) & &  \\
2011-01-13 & 55574 & bc196a(2) & 8.3 & 3.7 & 128 &  & C & 0.731 (0.089) & 0.39 (0.02) & 1.69 (0.22) & &  \\
 &  &  &  &  &  &  & J1 & 0.067 (0.016) & 1.40 (0.07) & & 2.06 (0.55) & $-$50.1 (1.3) \\
2011-06-28 & 55740 & rdv87 & 8.6 & 184 & 32 & \citep{2012AA...544A..34P} & C & 0.629 (0.082) & 0.33 (0.02) & 1.89 (0.27) & &  \\
 &  &  &  &  &  &  & J1 & 0.085 (0.018) & 1.48 (0.09) & & 1.57 (0.50) & $-$53.7 (1.3) \\
2012-02-13 & 55970 & bc201ae(2) & 8.3 & 3.2 & 128 &  & C & 0.696 (0.088) & 0.23 (0.01) & 4.63 (0.71) & &  \\
 &  &  &  &  &  &  & J1 & 0.077 (0.018) & 0.93 (0.05) & & 2.05 (0.54) & $-$54.0 (1.3) \\
2013-05-07 & 56419 & s4195a(2) & 7.6 & 1.2 & 256 &  & C & 0.498 (0.083) & 0.31 (0.03) & $>$1.77    & &  \\
 &  &  &  &  &  &  & J1 & 0.075 (0.022) & 0.52 (0.05) & & 2.36 (1.42) & $-$59.4 (1.0) \\
2013-06-22 & 56465 & s4195c & 7.6 & 0.8 & 256 &  & C & 0.451 (0.077) & 0.09 (0.01) & $>$1.59    & &  \\
 &  &  &  &  &  &  & J1 & 0.073 (0.021) & 0.31 (0.03) & & 2.33 (1.39) & $-$55.6 (1.0) \\
2013-12-11 & 56637 & rv102 & 8.6 & 63 & 32 &  & C & 0.526 (0.088) & 0.39 (0.04) & 1.12 (0.20) & &  \\
 &  &  &  &  &  &  & J1 & 0.079 (0.023) & 0.75 (0.07) & & 1.96 (0.70) & $-$73.9 (1.2) \\
2014-06-09 & 56817 & bg219d & 8.7 & 1 & 384 & \citep{2016AJ....151..154G} & C & 0.397 (0.058) & 0.47 (0.03) & 0.59 (0.09) & &  \\
 &  &  &  &  &  &  & J1 & 0.087 (0.014) & 1.34 (0.10) & & 2.42 (0.75) & $-$61.2 (1.3) \\
2014-08-05 & 56874 & bg219c & 8.7 & 1 & 384 & \citep{2016AJ....151..154G} & C & 0.449 (0.064) & 0.36 (0.02) & 1.13 (0.17) & &  \\
 &  &  &  &  &  &  & J1 & 0.080 (0.015) & 1.22 (0.08) & & 2.26 (0.96) & $-$66.6 (1.2) \\
2014-08-09 & 56878 & bg219e & 8.7 & 1 & 384 & \citep{2016AJ....151..154G} & C & 0.497 (0.086) & 0.38 (0.04) & 1.11 (0.20) & &  \\
 &  &  &  &  &  &  & J1 & 0.084 (0.024) & 0.97 (0.09) & & 2.14 (0.91) & $-$73.0 (1.2) \\
2015-06-11 & 57184 & bs241c(2) & 7.6 & 3 & 256 & \citep{2019MNRAS.485...88P} & C & 0.515 (0.066) & 0.34 (0.02) & 1.91 (0.27) & &  \\
 &  &  &  &  &  &  & J1 & 0.080 (0.014) & 1.18 (0.07) & & 2.32 (0.56) & $-$65.9 (1.3) \\
2015-09-05 & 57270 & bp192a7(2) & 7.6 & 2.7 & 256 & \citep{2021AJ....161...14P} & C & 0.282 (0.041) & 0.24 (0.02) & 2.01 (0.34) & &  \\
 &  &  &  &  &  &  & J1 & 0.066 (0.011) & 1.34 (0.10) & & 2.31 (1.00) & $-$71.5 (1.2) \\

%

2016-01-21 & 57408 & bp192e0(3) & 7.6 & 4.2 & 256 & \citep{2021AJ....161...14P} & C & 0.216 (0.033) & 0.32 (0.02) & 0.88 (0.14) & &  \\
 &  &  &  &  &  &  & J1 & 0.058 (0.009) & 1.62 (0.12) & & 2.45 (0.96) & $-$66.5 (1.2) \\
2016-05-30 & 57538 & bp192f9 & 7.6 & 1.7 & 256 & \citep{2021AJ....161...14P} & C & 0.195 (0.034) & 0.34 (0.03) & 0.69 (0.13) & &  \\
 &  &  &  &  &  &  & J1 & 0.048 (0.010) & 0.99 (0.10) & & 2.68 (0.56) & $-$60.3 (1.4) \\

 \bottomrule
\end{tabularx}

\end{table}

\begin{table}[H]\ContinuedFloat
\caption{{\em Cont.}}
    \small
\setlength{\cellWidtha}{\textwidth/13-2\tabcolsep+0.2in}
\setlength{\cellWidthb}{\textwidth/13-2\tabcolsep-0in}
\setlength{\cellWidthc}{\textwidth/13-2\tabcolsep-0in}
\setlength{\cellWidthd}{\textwidth/13-2\tabcolsep-0in}
\setlength{\cellWidthe}{\textwidth/13-2\tabcolsep-0in}
\setlength{\cellWidthf}{\textwidth/13-2\tabcolsep-0.2in}
\setlength{\cellWidthg}{\textwidth/13-2\tabcolsep-0in}
\setlength{\cellWidthh}{\textwidth/13-2\tabcolsep-0.2in}
\setlength{\cellWidthi}{\textwidth/13-2\tabcolsep+0.2in}
\setlength{\cellWidthj}{\textwidth/13-2\tabcolsep-0in}
\setlength{\cellWidthk}{\textwidth/13-2\tabcolsep-0in}
\setlength{\cellWidthl}{\textwidth/13-2\tabcolsep-0in}
\setlength{\cellWidthm}{\textwidth/13-2\tabcolsep-0in}

\begin{tabularx}{\textwidth}{
>{\centering\arraybackslash}m{\cellWidtha}
>{\centering\arraybackslash}m{\cellWidthb}
>{\centering\arraybackslash}m{\cellWidthc}
>{\centering\arraybackslash}m{\cellWidthd}
>{\centering\arraybackslash}m{\cellWidthe}
>{\centering\arraybackslash}m{\cellWidthf}
>{\centering\arraybackslash}m{\cellWidthg}
>{\centering\arraybackslash}m{\cellWidthh}
>{\centering\arraybackslash}m{\cellWidthi}
>{\centering\arraybackslash}m{\cellWidthj}
>{\centering\arraybackslash}m{\cellWidthk}
>{\centering\arraybackslash}m{\cellWidthl}
>{\centering\arraybackslash}m{\cellWidthm}
}

    \toprule
\multicolumn{2}{c}{\textbf{Observing Date}} 
& \textbf{Project} 
& \textbf{Frequency} 
& \textbf{On-Source} 
& \textbf{Bandwidth} 
& \multirow{2}{*}{\textbf{Ref.}} 
& \multirow{2}{*}{\textbf{Comp.}} 
& \textit{\textbf{S}} 
& \boldmath{$\phi$} 
& \textbf{\textit{T}\textsubscript{b}}
& \textit{\textbf{R}} 
& {\textbf{PA}} \\ 

\textbf{yyyy-mm-dd}
& \textbf{[MJD]} 
& \textbf{Code} 
& \textbf{[GHz] }
& \textbf{time} \textbf{[min]} 
& \textbf{[MHz]} &  &  
& \textbf{[Jy]} 
& \textbf{[mas]} 
&  \textbf{[10\textsuperscript{11}} \textbf{K]}
& \textbf{[mas]}
 & \textbf{[}\boldmath{$^{\circ}$}\textbf{]} \\
\midrule

 2017-01-17 & 57770 & uf001a & 8.7 & 0.2 & 384 &  & C & 0.396 (0.064) & 0.13 (0.01) & $>$3.62    & &  \\
 &  &  &  &  &  &  & J1 & 0.045 (0.017) & 1.15 (0.10) & & 2.41 (0.84) & $-$67.2 (1.2) \\

2017-01-22 & 57775 & uf001b & 8.7 & 85 & 384 &  & C & 0.429 (0.061) & 0.15 (0.01) & 5.89 (1.14) & &  \\
 &  &  &  &  &  &  & J1 & 0.052 (0.015) & 0.77 (0.05) & & 2.18 (0.54) & $-$72.9 (1.3) \\
 &  &  &  &  &  &  & J3 & 0.008 (0.006) & 0.19 (0.01) & & 4.48 (0.54) & $-$56.7 (1.5) \\
2017-02-19 & 57803 & uf001c & 8.7 & 21.6 & 384 &  & C & 0.441 (0.062) & 0.26 (0.02) & 2.18 (0.35) & &  \\
 &  &  &  &  &  &  & J1 & 0.063 (0.015) & 0.83 (0.06) & & 2.07 (0.50) & $-$76.8 (1.3) \\
 &  &  &  &  &  &  & J3 & 0.011 (0.006) & 0.43 (0.03) & & 4.52 (0.50) & $-$55.9 (1.5) \\
2017-02-24 & 57808 & uf001d & 8.7 & 0.6 & 384 &  & C & 0.478 (0.071) & 0.28 (0.02) & 1.95 (0.32) & &  \\
 &  &  &  &  &  &  & J1 & 0.080 (0.018) & 1.04 (0.08) & & 2.16 (0.55) & $-$74.6 (1.3) \\
2017-03-27 & 57839 & uf001f & 8.7 & 44 & 384 &  & C & 0.487 (0.072) & 0.31 (0.02) & 1.69 (0.27) & &  \\
 &  &  &  &  &  &  & J1 & 0.067 (0.018) & 0.74 (0.06) & & 2.26 (0.48) & $-$71.5 (1.4) \\
2017-04-28 & 57871 & uf001g & 8.7 & 0.5 & 384 &  & C & 0.436 (0.072) & 0.23 (0.02) & 2.63 (0.49) & &  \\
 &  &  &  &  &  &  & J1 & 0.059 (0.019) & 0.78 (0.07) & & 2.30 (0.48) & $-$67.4 (1.4) \\
2017-05-01 & 57874 & uf001h & 8.7 & 0.3 & 384 &  & C & 0.488 (0.079) & 0.31 (0.03) & 1.70 (0.30) & &  \\
 &  &  &  &  &  &  & J1 & 0.048 (0.021) & 0.77 (0.07) & & 2.03 (0.96) & $-$72.3 (1.1) \\
2017-06-10 & 57914 & uf001k & 8.7 & 0.5 & 384 &  & C & 0.597 (0.072) & 1.00 (0.05) & 0.19 (0.02) & &  \\
 &  &  &  &  &  &  & J2 & 0.041 (0.027) & 0.52 (0.02) & & 3.48 (2.61) & $-$82.5 (0.9) \\
2017-06-16 & 57920 & uf001l & 8.7 & 0.3 & 384 &  & C & 0.587 (0.081) & 0.30 (0.02) & 2.15 (0.33) & &  \\
 &  &  &  &  &  &  & J1 & 0.063 (0.019) & 0.61 (0.04) & & 2.14 (0.45) & $-$70.0 (1.4) \\
 &  &  &  &  &  &  & J3 & 0.015 (0.007) & 0.13 (0.01) & & 4.37 (0.45) & $-$51.8 (1.5) \\
2017-07-09 & 57943 & uf001m & 8.7 & 0.8 & 384 &  & C & 0.650 (0.087) & 0.30 (0.02) & 2.31 (0.35) & &  \\
 &  &  &  &  &  &  & J1 & 0.069 (0.019) & 0.50 (0.03) & & 2.10 (0.46) & $-$74.0 (1.4) \\
 &  &  &  &  &  &  & J3 & 0.018 (0.007) & 0.53 (0.03) & & 4.55 (0.46) & $-$55.5 (1.5) \\
2017-08-05 & 57970 & uf001o & 8.7 & 47 & 384 &  & C & 0.601 (0.083) & 0.19 (0.01) & 5.15 (0.89) & &  \\
 &  &  &  &  &  &  & J1 & 0.045 (0.019) & 1.07 (0.07) & & 2.35 (0.94) & $-$74.7 (1.2) \\
2017-08-13 & 57978 & uf001p & 8.7 & 0.3 & 384 &  & C & 0.603 (0.078) & 0.34 (0.02) & 1.65 (0.23) & &  \\
 &  &  &  &  &  &  & J3 & 0.019 (0.006) & 0.32 (0.02) & & 4.25 (0.97) & $-$71.7 (1.3) \\
2017-10-22 & 58048 & uf001t & 8.7 & 37 & 384 &  & C & 0.608 (0.082) & 0.29 (0.02) & 2.29 (0.35) & &  \\
 &  &  &  &  &  &  & J1 & 0.059 (0.019) & 0.39 (0.02) & & 1.89 (1.04) & $-$80.3 (1.1) \\
 &  &  &  &  &  &  & J2 & 0.033 (0.018) & 1.63 (0.10) & & 3.39 (1.04) & $-$58.5 (1.3)\\

2018-01-19 & 58137 & ug002a(6) & 8.7 & 3.7 & 384 &  & C & 0.636 (0.084) & 0.42 (0.03) & 1.18 (0.17) & &  \\
 &  &  &  &  &  &  & J1 & 0.072 (0.018) & 0.67 (0.04) & & 2.16 (0.49) & $-$75.3 (1.3) \\
 &  &  &  &  &  &  & J3 & 0.012 (0.007) & 0.34 (0.02) & & 4.29 (0.49) & $-$60.7 (1.5) \\
2018-08-10 & 58340 & ug002o(6) & 8.7 & 4.5 & 384 &  & C & 0.439 (0.056) & 0.33 (0.02) & 1.32 (0.19) & &  \\
 &  &  &  &  &  &  & J1 & 0.035 (0.011) & 0.70 (0.04) & & 2.18 (0.54) & $-$82.7 (1.3) \\
 &  &  &  &  &  &  & J2 & 0.027 (0.004) & 2.03 (0.11) & & 3.35 (0.54) & $-$61.8 (1.4) \\

  \bottomrule
\end{tabularx}

\end{table}

\begin{table}[H]\ContinuedFloat
\caption{{\em Cont.}}
    \small
\setlength{\cellWidtha}{\textwidth/13-2\tabcolsep+0.2in}
\setlength{\cellWidthb}{\textwidth/13-2\tabcolsep-0in}
\setlength{\cellWidthc}{\textwidth/13-2\tabcolsep-0in}
\setlength{\cellWidthd}{\textwidth/13-2\tabcolsep-0in}
\setlength{\cellWidthe}{\textwidth/13-2\tabcolsep-0in}
\setlength{\cellWidthf}{\textwidth/13-2\tabcolsep-0.2in}
\setlength{\cellWidthg}{\textwidth/13-2\tabcolsep-0in}
\setlength{\cellWidthh}{\textwidth/13-2\tabcolsep-0.2in}
\setlength{\cellWidthi}{\textwidth/13-2\tabcolsep+0.2in}
\setlength{\cellWidthj}{\textwidth/13-2\tabcolsep-0in}
\setlength{\cellWidthk}{\textwidth/13-2\tabcolsep-0in}
\setlength{\cellWidthl}{\textwidth/13-2\tabcolsep-0in}
\setlength{\cellWidthm}{\textwidth/13-2\tabcolsep-0in}

\begin{tabularx}{\textwidth}{
>{\centering\arraybackslash}m{\cellWidtha}
>{\centering\arraybackslash}m{\cellWidthb}
>{\centering\arraybackslash}m{\cellWidthc}
>{\centering\arraybackslash}m{\cellWidthd}
>{\centering\arraybackslash}m{\cellWidthe}
>{\centering\arraybackslash}m{\cellWidthf}
>{\centering\arraybackslash}m{\cellWidthg}
>{\centering\arraybackslash}m{\cellWidthh}
>{\centering\arraybackslash}m{\cellWidthi}
>{\centering\arraybackslash}m{\cellWidthj}
>{\centering\arraybackslash}m{\cellWidthk}
>{\centering\arraybackslash}m{\cellWidthl}
>{\centering\arraybackslash}m{\cellWidthm}
}

    \toprule
\multicolumn{2}{c}{\textbf{Observing Date}} 
& \textbf{Project} 
& \textbf{Frequency} 
& \textbf{On-Source} 
& \textbf{Bandwidth} 
& \multirow{2}{*}{\textbf{Ref.}} 
& \multirow{2}{*}{\textbf{Comp.}} 
& \textit{\textbf{S}} 
& \boldmath{$\phi$} 
& \textbf{\textit{T}\textsubscript{b}}
& \textit{\textbf{R}} 
& {\textbf{PA}} \\ 

\textbf{yyyy-mm-dd} 
& \textbf{[MJD]} 
& \textbf{Code} 
& \textbf{[GHz] }
& \textbf{time} \textbf{[min]} 
& \textbf{[MHz]} &  &  
& \textbf{[Jy]} 
& \textbf{[mas]} 
&  \textbf{[10\textsuperscript{11}} \textbf{K]}
& \textbf{[mas]}
 & \textbf{[}\boldmath{$^{\circ}$}\textbf{]} \\
\midrule

 2019-08-05 & 58700 & bl273a & 15.0 & 35 & 256 & \citep{2009AJ....137.3718L} & C & 1.204 (0.134) & 0.10 (0.01) & 13.47 (3.13) & &  \\
 &  &  &  &  &  &  & J1 & 0.051 (0.039) & 3.15 (0.10) & & 2.34 (0.29) & $-$74.9 (1.4) \\

2019-12-31 & 58848 & bl229nd & 15.0 & 35 & 256 & \citep{2009AJ....137.3718L} & C & 1.217 (0.203) & 0.05 (0.01) & $>$26.94    & &  \\
 &  &  &  &  &  &  & J1 & 0.321 (0.118) & 2.50 (0.23) & & 2.23 (0.31) & $-$73.0 (1.4) \\
2020-05-25 & 58994 & bl229be & 15.0 & 35 & 256 & \citep{2009AJ....137.3718L} & C & 2.139 (0.225) & 0.02 (0.01) & $>$159.00    & &  \\
 &  &  &  &  &  &  & J1 & 0.039 (0.024) & 2.31 (0.05) & & 2.32 (0.64) & $-$78.6 (1.3) \\
2020-06-08 & 59008 & bl273f & 15.0 & 44 & 256 & \citep{2009AJ....137.3718L} & C & 2.169 (0.230) & 0.06 (0.01) & 77.37 (29.29) & &  \\
 &  &  &  &  &  &  & J1 & 0.037 (0.025) & 1.97 (0.05) & & 2.21 (0.31) & $-$76.4 (1.4) \\
2020-09-24 & 59116 & br235b & 7.6 & 0.8 & 256 &  & C & 0.846 (0.111) & 0.02 (0.01) & $>$65.13    & &  \\
 &  &  &  &  &  &  & J2 & 0.032 (0.010) & 1.67 (0.10) & & 3.54 (0.51) & $-$60.9 (1.4) \\
2020-10-21 & 59143 & bl273g1 & 15.0 & 44 & 256 & \citep{2009AJ....137.3718L} & C & 1.717 (0.192) & 0.09 (0.01) & 20.61 (4.92) & &  \\
 &  &  &  &  &  &  & J1 & 0.032 (0.029) & 0.53 (0.02) & & 2.09 (0.29) & $-$80.2 (1.4) \\
2020-11-21 & 59174 & bl273h & 15.0 & 44 & 256 & \citep{2009AJ....137.3718L} & C & 2.298 (0.246) & 0.13 (0.01) & 13.71 (2.51) & &  \\
 &  &  &  &  &  &  & J1 & 0.059 (0.029) & 1.72 (0.04) & & 2.03 (0.34) & $-$80.9 (1.4) \\
2020-12-24 & 59207 & bl229bl & 15.0 & 35 & 256 & \citep{2009AJ....137.3718L} & C & 1.896 (0.213) & 0.10 (0.01) & 19.19 (4.29) & &  \\
 &  &  &  &  &  &  & J1 & 0.053 (0.032) & 0.70 (0.02) & & 2.01 (0.29) & $-$83.3 (1.4) \\
2021-01-10 & 59224 & bl273i & 15.0 & 44 & 256 & \citep{2009AJ....137.3718L} & C & 2.463 (0.269) & 0.10 (0.01) & 27.76 (6.43) & &  \\
 &  &  &  &  &  &  & J1 & 0.055 (0.036) & 0.58 (0.02) & & 2.01 (0.32) & $-$82.3 (1.4) \\
2021-03-07 & 59280 & uh007g & 8.7 & 2.5 & 512 &  & C & 1.878 (0.217) & 0.12 (0.01) & 39.36 (7.79) & &  \\
 &  &  &  &  &  &  & J1 & 0.124 (0.036) & 0.69 (0.03) & & 1.94 (0.56) & $-$85.2 (1.3) \\
2021-10-18 & 59505 & bp252i & 7.6 & 1.1 & 512 &  & C & 1.448 (0.189) & 0.17 (0.01) & 20.17 (3.51) & &  \\
 &  &  &  &  &  &  & J2 & 0.036 (0.016) & 1.19 (0.07) & & 4.19 (0.51) & $-$48.8 (1.4) \\
2021-11-04 & 59522 & uh007o & 8.7 & 0.02 & 384 &  & C & 2.227 (0.274) & 0.32 (0.02) & 6.92 (0.95) & &  \\
 &  &  &  &  &  &  & J1 & 0.066 (0.052) & 0.57 (0.03) & & 2.26 (0.90) & $-$73.2 (1.2) \\
2022-10-16 & 59868 & uh007y(4) & 8.7 & 17.7 & 384 &  & C & 1.026 (0.123) & 0.20 (0.01) & 8.58 (1.35) & &  \\
 &  &  &  &  &  &  & J1 & 0.059 (0.023) & 1.00 (0.05) & & 2.25 (0.52) & $-$84.7 (1.3)\\
 \bottomrule
   \end{tabularx}
   
    \label{tab:results}

\begin{adjustwidth}{+\extralength}{0cm}
    \noindent{\footnotesize{Notes: Col.~1---observing date; Col.~2---observing date in MJD; Col.~3---project code; Col.~4---observing frequency; Col.~5---sum of the scan durations; Col.~6---total observing bandwidth; \mbox{Col.~7---literature} reference; Col.~8---jet component designation; Col.~9---fitted flux density of the component; Col.~10---fitted FWHM of the circular Gaussian component; Col.~11---redshift-corrected core brightness temperature; Col.~12---angular separation of the jet component from the core; Col.~13---position angle of the component with respect to the core, measured in degrees from north to east. Uncertainties are given in parentheses.}}
\end{adjustwidth}
\end{table}
\finishlandscape

\begin{adjustwidth}{-\extralength}{0cm}
\printendnotes[custom] 

\reftitle{References}



\PublishersNote{}

\end{adjustwidth}
\end{document}